\documentclass[12pt]{article}
\usepackage{amsthm,amssymb,amsmath}
\usepackage[british]{babel}

 1 
1

\newcommand{\n}{\ensuremath{\mathfrak{N}}}

\newcommand{\bt}{{\boldsymbol{t}}}

\newcommand{\bu}{{\boldsymbol{u}}}

\renewcommand{\d}{\operatorname{d}}

\newtheorem{pro}{Proposition}

\newcommand{\be}{\begin{equation}}
\newcommand{\ee}{\end{equation}}
\begin{document}

\title{\sc On the Whitham Hierarchies:
Reductions and Hodograph Solutions\thanks{Partially supported by
CICYT proyecto PB98--0821 }}
\author{Francisco Guil, Manuel Ma\~{n}as$^{\dag}$ and
Luis Mart\'{\i}nez Alonso$^{\ddag}$\\
\emph{Departamento de F\'{\i}sica Te\'{o}rica II, Universidad
Complutense}\\ \emph{E28040 Madrid, Spain} \\
\texttt{$^\dag$manuel@darboux.fis.ucm.es}\\
\texttt{$^\ddag$luism@fis.ucm.es}\\
}
\date{} \maketitle
\begin{abstract}
A general scheme for analyzing reductions of Whitham hierarchies
is presented. It is based on a method for determining the
$S$-function by means of a system of first order partial
differential equations. Compatibility systems of differential
equations characterizing both reductions and hodograph solutions
of Whitham hierarchies are obtained. The method is illustrated by
exhibiting solutions of integrable models such as the
dispersionless Toda  equation (heavenly equation)  and the
generalized Benney system.
\end{abstract}

\vspace*{.5cm}

\begin{center}\begin{minipage}{12cm}

\end{minipage}
\end{center}
\newpage

\section{Introduction}

The study of dispersionless (or quasiclasssical) limits of
integrable systems of KdV-type and their applications has been an
active subject of research for more than twenty years (see for
example \cite{1}-\cite{14}). However, despite of the fact that
many important developments on the algebraic and geometric aspects
of these systems have been made,  the theory of their solution
methods seems far from being completed. Indeed, only for a few
cases \cite{15}-\cite{17} the dispersionless limit of the inverse
scattering method is available and \emph{dispersionless versions}
of ordinary direct methods like the $\overline{\partial}$-method
are not yet fully developed \cite{18}.

 In \cite{3}-\cite{4} Kodama and Gibbons provided a direct method
for finding solutions of the dispersionless KP (dKP) equation
\begin{equation}\label{dkpe}
(u_t+3uu_x)_x=\frac{3}{4}u_{yy},
\end{equation}
and its associated dKP hierarchy of nonlinear systems. The main
ingredient of their method is the use of reductions of the dKP
hierarchy formulated in terms of hydrodynamic-type equations. As a
consequence it follows that solutions of the dKP hierarchy turn
out to be determined through hodograph equations. Recently, we
proposed \cite{19} an alternative direct method for solving the
dKP hierarchy from its reductions. It is based on the
characterization of reductions and hodograph solutions of the dKP
hierarchy by means of certain systems of first-order partial
differential equations.

The aim of this paper is to  present a generalization of the
method of \cite{19} which applies to the Whitham hierachies of
dispersionless integrable systems. These hierarchies were
introduced by Krichever in \cite{7} and  contain many interesting
dispersionless models as, for example, the $(2+1)$-dimensional
integrable systems
\begin{equation}\label{2T}
\Phi_{xy}+\Big(e^{\Phi}\Big)_{tt}=0,
\end{equation}
known as the dispersionless Toda (dT) equation (\emph{heavenly
equation} or Boyer-Finley equation \cite{20}-\cite{21}), and the
generalized Benney system \cite{10}
\begin{equation}\label{bs}\begin{gathered}
a_t+(av)_t=0,\\
v_t+vv_x+w_x=0,\\
w_y+a_x=0,
\end{gathered}
\end{equation}

In the next section we review briefly the definition of the
Whitham hierarchies (zero genus case) and introduce our main
notation conventions. Section 3 concerns with the  method for
characterizing reductions and hodograph solutions of the Whitham
hierarchies. To this end we take advantage of the same scheme as
in \cite{19} to introduce reductions through systems of
first-order partial differential equations. The main difference
with respect to the procedure used in \cite{19} lies in the more
involved construction of the $S$-function. Like in the study of
the dKP hierarchy, we find that the compatibility equations for
characterizing diagonal reductions of the Whitham hierarchies are
deeply connected with the theory of Combescure transformations of
conjugate nets. Finally, Section 4 is devoted to illustrate the
method with examples of hodograph solutions of \eqref{2T} and
\eqref{bs}.

\section{The Whitham hierarchy}

The $M$-th Whitham hierarchy is related to a family of evolution
equations for a set of $M$ functions
$z_{\alpha}=z_{\alpha}(p,\bt),\; 1\leq \alpha\leq M$  depending on
a complex variable $p$ and an infinite set of complex time
parameters
\[
\bt:=\{t_A:A=(\alpha,n)\in\boldsymbol{A}\},
\]
where
\[
 \boldsymbol{A}=\{(\alpha,0)\}_{\alpha=2}^M\bigcup\{(\alpha,n)\}_{\substack{\alpha=1,\dots,M\\n=1,\dots,\infty}}.
\]
It is assumed that a neighborhood $\mathcal{D}$ of $\infty$ in the
extended complex plane of the $p$ variable exists on which each
$z_{\alpha}$ has a simple pole at an associated point
$q_{\alpha}=q_{\alpha}(\bt)$ . In particular, we set $q_1=\infty$
and assume that $z_1$ posses the normalized Laurent expansion
\begin{equation}\label{1}
z_1(p,\bt)=p+\sum_{n=1}^{\infty}\frac{a_{1,n}(\bt)}{p^n},\quad
p\rightarrow\infty.
\end{equation}
The corresponding expansions for the remaining functions
$z_{\alpha}$ at $q_{\alpha}$ will be written as
\begin{equation}\label{2}
z_i(p,\bt)=\frac{a_{i,-1}(\bt)}{p-q_i(\bt)}+\sum_{n=0}^{\infty}a_{i,n}
(\bt)\big(p-q_i(\bt)\big)^n,\quad p\rightarrow q_i(\bt),\quad
2\leq i\leq M.
\end{equation}
In order to define the Whitham equations we introduce the system
of evolution equations
\begin{equation}\label{3}
 \frac{\partial z_{\alpha}}{\partial t_A}=\{\Omega_A,z_{\alpha}\},\quad
 1\leq\alpha\leq M.
\end{equation}
Here $\{\cdot,\cdot\}$ is the Poisson bracket
\[
\{F_1,F_2\}:=\frac{\partial F_1}{\partial p}\frac{\partial
F_2}{\partial x}-\frac{\partial F_1}{\partial x}\frac{\partial
F_2}{\partial p},\quad x:=t_{1,1},
\]
and the functions $\Omega_A=\Omega_A(p,\bt)$ are defined by
\begin{equation}\label{4}
\Omega_A=\left\{\begin{array}{ll} -\ln(p-q_i(\bt))&\mbox{for
$A=(i,0),\;2\leq i\leq M$,}\\\\
\big((z_{\alpha})^n\big)_+ &\mbox{for $A=(\alpha,n),\; 1\leq
\alpha\leq M, \;n\geq 1 $,}
\end{array}
\right.
\end{equation}
where
\[
\big((z_{\alpha})^n\big)_+:=P_{(\alpha,+)}(z_{\alpha}^n),
\]
with $P_{(\alpha,+)}$ being the following projectors acting on
Laurent series around $p=q_{\alpha}(\bt)$
\[\begin{gathered}
P_{(1,+)}\big(\sum_{n=-\infty}^{\infty}a_np^n\big)=\sum_{n=0
}^{\infty}a_np^n,\\
P_{(i,+)}\big(\sum_{n=-\infty}^{\infty}b_n
\big(p-q_i(\bt)\big)^n\big)=\sum_{n=1}^{\infty}\frac{b_{-n}}
{(p-q_i(\bt))^n},\quad 2\leq i\leq M.
\end{gathered}
\]

The Whitham hierarchy is the set of equations
\begin{equation}\label{5}
\frac{\partial \Omega_A}{\partial t_B}-\frac{\partial
\Omega_B}{\partial t_A}+\{\Omega_A, \Omega_B \}=0,\quad
A,B\in\boldsymbol{A},
\end{equation}
which describe the compatibility conditions for the system
\eqref{3}. For $M=1$ the Whitham hierarchy becomes the
dispersionless Kadomtsev--Petviasvhili  (dKP) hierarchy. Some
interesting nonlinear models included  in the case $M=2$ are, for
example,
\begin{description}
\item{1.} The dispersionless Toda (dT) equation (\emph{heavenly
equation} or Boyer--Finley equation)

\begin{equation}\label{6}
\Phi_{xy}+(\exp(\Phi))_{tt}=0,
\end{equation}
which is obtained from \eqref{5} by setting $A=(2,0)$, $B=(2,1)$
and
\begin{equation}\label{7}\begin{gathered}
\;y:=t_{(2,1)},\; t:=-t_{(2,0)},\\
\Phi:=\ln a_{2,-1},\\
\Omega_{(2,0)}=-\ln(p-q_2),\quad
\Omega_{(2,1)}=\frac{a_{2,-1}(\bt)}{p-q_2(\bt)}.
\end{gathered}
\end{equation}
\item{2.} The generalized  Benney system (generalized gas
equation) \cite{10}
\begin{equation}\label{8}\begin{gathered}
a_t+(av)_x=0,\\
v_t+vv_x+w_x=0,\\
w_y+a_x=0,
\end{gathered}
\end{equation}
can be regarded as a two-dimensional generalization of the
equations for one-dimensional gas dynamics. It takes the form
\eqref{5} by setting $A=(2,1)$, $B=(1,2)$ and
\begin{equation}\label{9}\begin{gathered}
y:=t_{(2,1)},\; t:=-\frac{1}{2}t_{(1,2)},\\
a:=a_{2,-1},\;\; v:=q_2,\;\; w:=a_{1,1}\\
\Omega_{(2,1)}=\frac{a_{2,-1}(\bt)}{p-q_2(\bt)},\;\;
\Omega_{(1,2)}=p^2+2a_{1,1}.
\end{gathered}
\end{equation}
\end{description}

\section{Reductions of the Whitham hierarchy}

\subsection{The $S$ function}

In this paper se shall study \emph{algebraic orbits} of the zero
genus Whitham hierarchy defined by \cite{8}
\begin{equation}\label{10}
z_i=f_{i}(z),\quad z:=z_1,\quad 2\leq i\leq M,
\end{equation}
which are easily checked to be compatible with \eqref{3}.

 Furthermore, it follows from \eqref{3} and
\eqref{5} that
\[
\frac{\partial}{\partial
t_B}\Omega_A(p(z,\bt),\bt)=\frac{\partial}{\partial
t_A}\Omega_B(p(z,\bt),\bt),
\]
and therefore there exists a potential function $S=S(z,\bt)$
satisfying
\begin{equation}\label{11}
\frac{\partial S(z,\bt)}{\partial
t_A}=\Omega_A(p(z,\bt),\bt),\quad A\in\boldsymbol{A}.
\end{equation}

Reciprocally, we can state the following proposition on which our
solution method for the Whitham hierarchy will be based
\begin{pro}
Let $\{z_{\alpha}(p,\bt)\}_{\alpha =1}^M$ be a set of functions
satisfying a system of time-independent relations \eqref{10} as
well as \eqref{1} and \eqref{2}. If a function $S(z,\bt)$
verifying \eqref{11} exists, then the functions
$z_{\alpha}(p,\bt)$ provide a solution of the Whitham hierarchy.
\end{pro}
\begin{proof}

First we notice that by setting $A=(1,1)$ in \eqref{11} it follows
\[
p(z,\bt)=\frac{\partial S(z,\bt)}{\partial x},
\]
so that
\begin{align*}
\frac{\partial p}{\partial t_A}&=\frac{\partial }{\partial
x}\frac{\partial S(z,\bt)}{\partial t_A}=\frac{\partial
}{\partial x}\Omega_A(p(z,\bt),\bt)\\
&=\frac{\partial \Omega_A}{\partial p}\frac{\partial p}{\partial
x}+\frac{\partial \Omega_A}{\partial x}.
\end{align*}
Hence, the function $z=z(p,\bt)$ satisfies
\begin{align*}
\frac{\partial z}{\partial t_A}&=-\frac{\partial z}{\partial
p}\frac{\partial p}{\partial t_A}=-\frac{\partial z}{\partial
p}\Big(\frac{\partial \Omega_A}{\partial p}\frac{\partial
p}{\partial x}+\frac{\partial
\Omega_A}{\partial x}\Big)\\
&=\frac{\partial\Omega_A}{\partial p}\frac{\partial z}{\partial x}
-\frac{\partial\Omega_A}{\partial x}\frac{\partial z}{\partial p}
=\{\Omega_A,z\}.
\end{align*}
Therefore, by using \eqref{10} we deduce \eqref{3}.
\end{proof}

\subsection{$N$-reductions of the Whitham hierarchy}

We are going to describe a method for finding solutions of the
Whitham  hierarchy from  functions $z=z(p,\bu)$ depending on $p$
and a finite set of variables $\bu:=(u_1,\ldots,u_N)$, such that
the inverse function $p=p(z,\bu)$ satisfies a system of equations
of the form
\begin{equation}\label{14}
\frac{\partial p}{\partial u_i}=R_i(p,\bu),\quad 1\leq i\leq N,
\end{equation}
or, equivalently, in terms of $z=z(p,\bu)$
\begin{equation}\label{13}
\frac{\partial z}{\partial u_i}+R_i(p,\bu)\frac{\partial
z}{\partial p}=0,\quad 1\leq i\leq N.
\end{equation}
The following conditions for the functions $R_i$ will be  assumed
\begin{enumerate}
\item[i)] The functions $R_i$ are rational functions of $p$ which
have singularities only at $N$ simple poles $p_i= p_i(\bu),\;
i=1,\ldots ,N$, and vanish at $p=\infty$. Therefore, they can be
expanded as
\begin{equation}\label{15}
R_i(p,\bu)=\sum_{j=1}^N \frac{r_{ij}(\bu)}{p-p_j(\bu)}.
\end{equation}

\item[ii)] The functions $r_{ij}(\bu),\; p_i(\bu)$ satisfy the
compatibility conditions for \eqref{13}-\eqref{14}
\begin{equation}\label{16}
\begin{gathered}
r_{ik}\dfrac{\partial p_k}{\partial u_j}-r_{jk}\dfrac{\partial
p_k}{\partial u_i}=\sum_{l\neq k}
\frac{r_{jl}r_{ik}-r_{il}r_{jk}}{p_k-p_l},\\\\
\dfrac{\partial r_{ik}}{\partial u_j}-\dfrac{\partial
r_{jk}}{\partial u_i} =2\sum_{l\neq k}
\frac{r_{jk}r_{il}-r_{ik}r_{jl}}{(p_k-p_l)^2},
\end{gathered}
\end{equation}
where $i\neq j$.
\end{enumerate}

The starting point of the method is a solution $z=z(p,\bu)$  of
\eqref{14} with a Laurent expansion
\begin{equation}\label{12}
z(p,\bu)=p+\sum_{n=1}^{\infty}\frac{a_n(\bu)}{p^n},\quad
p\rightarrow\infty,
\end{equation}
which is assumed to define a univalent analytic function
$z:\mathcal{D}\rightarrow \mathcal{D}'$ between two neighborhoods
$\mathcal{D}$ and $\mathcal{D}'$ of $\infty$ in the extended
complex planes of the variables $p$ and $z$ respectively. The next
step is to take $(M-1)$ different points $z_{0,i}\in
\mathcal{D}',\;2\leq i\leq M$ and define the functions
\begin{equation}\label{17}
\begin{gathered}
z_1(p,\bu):=z(p,\bu),\\
z_i(p,\bu):=\frac{1}{z(p,u)-z_{0,i}},\quad 2\leq i\leq M.
\end{gathered}
\end{equation}
Obviously, they satisfy the system of equations
\begin{equation}\label{17'}
\frac{\partial z_{\alpha}}{\partial u_i}+R_i(p,\bu)\frac{\partial
z_{\alpha}}{\partial p}=0,\quad 1\leq \alpha \leq M,
\end{equation}
and admit expansions of the form
\begin{equation}\label{18}
\begin{gathered}
z_1(p,\bu)=p+\sum_{n=1}^{\infty}\frac{a_{1,n}(\bu)}{p^n},\quad
p\rightarrow\infty.\\
z_i(p,\bu)=\frac{a_{i,-1}(\bu)}{p-q_i(\bu)}+\sum_{n=0}^{\infty}a_{i,n}
(\bu)\big(p-q_i(\bu)\big)^n,\quad p\rightarrow q_i(\bu),
\end{gathered}
\end{equation}
for $2\leq i\leq M$, here
\begin{equation}\label{19}
q_i(\bu):=p(z_{0,i},\bu).
\end{equation}

Observe that introducing the expansions at $p=\infty,q_i$,
$i=2,\dots,M$, of
 \eqref{18} in \eqref{17'} we get
\begin{align}
&\label{poten0}\begin{cases}
\begin{aligned}
\dfrac{\partial a_{1,1}}{\partial u_i}&=-\sum_{j=1}^Nr_{ij},\\
\dfrac{\partial a_{1,2}}{\partial u_i}&=-\sum_{j=1}^Nr_{ij}p_j,\\
\dfrac{\partial (a_{1,3}+a_{1,1}^2/2)}{\partial u_i}&=
-\sum_{j=1}^Nr_{ij}p_j^2,
\end{aligned}
\end{cases}
\\
\label{poten1} & \begin{cases}\begin{aligned}
\frac{\partial q_\alpha }{\partial u_j}&=R_j(q_\alpha ),\\
\frac{\partial \log a_{\alpha ,-1}}{\partial u_j}&=\frac{\d
R_j}{\d p}(q_\alpha )
\end{aligned}
\end{cases} \text{for $\alpha =2,\dots,M$}
\end{align}
while the other coefficients $a_{\alpha ,n}$ in the expansion of
$z_\alpha $ are determined by:
\[
\begin{aligned}
\frac{\partial a_{1,n}}{\partial u_j}&=-R_{j,n}+\sum_{k=1}^{n-2}
(n-k)R_{j,k}a_{i,n-k},\\
 \frac{\partial
a_{i,n}}{\partial u_j}&=\sum_{k=1}^{n+2}\frac{1}{k!}\frac{\d^{k}
R_j}{\d p^{k}}(q_i)a_{i,n-k+1},\quad  \text{for $i=2,\dots,M$},
\end{aligned}
\]
with
\[
R_{j,k}=\sum_{i=1}^Nr_{ji}p_i^{k-1}.
\]

Finally, we introduce the function
\begin{equation}\label{20}
\begin{gathered}
\mathcal{S}(p,\bu,\bt)=\mathcal{S}_+(p,\bu,\bt)+\mathcal{S}_-(p,\bu),
\\
\mathcal{S}_+:=\sum_{A\in\boldsymbol{A}}t_A\Omega_A(p,\bu),
\end{gathered}
\end{equation}
where $\Omega_A(p,\bu)$ are defined by \eqref{4} and \eqref{18},
and $\mathcal{S}_-(p,\bu)$ is  an analytic function on $
\mathcal{D}$ such that
\begin{equation}\label{21}
\lim_{p\rightarrow\infty} \mathcal{S}_-(p,\bu)=0.
\end{equation}

We can now enounce the following statement
\begin{pro}
If $\mathcal{S}_-(p,\bu)$ satisfies a system of equations
\begin{equation}\label{22}
\frac{\partial \mathcal{S}_-}{\partial u_i}+R_i\frac{\partial
\mathcal{S}_-}{\partial p}=\sum_k \frac{r_{ik}F_k}{p-p_k},\quad
1\leq i\leq N,
\end{equation}
for a given set of functions $\{F_i=F_i(\bu)\}_{i=1}^N$ verifying
the compatibility conditions for \eqref{22}
\begin{equation}\label{23}
r_{ik}\dfrac{\partial F_k}{\partial u_j}-r_{jk}\dfrac{\partial
F_k}{\partial u_i}=\sum_{l\neq k}
\frac{r_{jl}r_{ik}-r_{il}r_{jk}}{(p_k-p_l)^2}(F_k-F_l),\quad i\neq
j,
\end{equation}
and the functions  $\{u_i=u_i(\bt)\}_{i=1}^N$ are implicitly
determined by means of the \emph{hodograph relations}
\begin{equation}\label{24}
\sum_{A\in\boldsymbol{A}}t_A\frac{\partial\Omega_A}{\partial p}
(p_i(\bu),\bu)+F_i(\bu)=0,\quad 1\leq i\leq N,
\end{equation}
then
\begin{equation}\label{25}
S(z,\bt):=\mathcal{S}\big(p(z,\bu(\bt)),\bu(\bt),\bt\big),
\end{equation}
is an $S$-function for the Whitham hierarchy.
\end{pro}

\begin{proof}
The proof is based on the following consequence of \eqref{20} and
\eqref{25}
\begin{equation}\label{26}
\frac{\partial}{\partial
t_A}S(z,\bt)=\Omega_A(p(z,\bu(\bt)),\bt)+\sum_{i=1}^N\frac{\partial
u_i}{\partial t_A}\Big(\frac{\partial}{\partial
u_i}\mathcal{S}(p(z,\bu),\bu,\bt)\Big)\Big|_{\bu=\bu(\bt)},
\end{equation}
and our aim is to prove that under the hypothesis of the
proposition the functions
\begin{equation}\label{27}
\frac{\partial}{\partial
u_i}\mathcal{S}(p(z,\bu),\bu,\bt)=\frac{\partial\mathcal{S}}{\partial
p}R_i+\frac{\partial\mathcal{S}}{\partial u_i}.
\end{equation}
vanish identically, so that $S(z,\bt)$ satisfies \eqref{11} and,
consequently,  the statement will follow at once.

By construction the functions \eqref{27} are analytic on
$\mathcal{D}$ up to a set of possible isolated singularities at
$\{p_i(\bu),q_{\alpha}(\bu)\}$. On the other hand we observe that
\eqref{22} implies
\begin{equation}\label{28}
F_i(\bu)=\frac{\partial \mathcal{S}_-}{\partial p}(p_i(\bu),\bu),
\end{equation}
so that \eqref{24} is equivalent to
\begin{equation}\label{29}
\frac{\partial\mathcal{S}}{\partial p}(p_i(\bu),\bu)=0,\quad 1\leq
i\leq N.
\end{equation}
As a consequence we deduce that the functions \eqref{27} are
analytic at $p_i(\bu)$.  Hence, their possible singularities
reduce to the points $q_{\alpha}(\bu)$. However, we have
\begin{equation}\label{30}
\frac{\partial}{\partial u_i}\mathcal{S}(p(z,\bu),\bu,\bt)=
\sum_{A\in\boldsymbol{A}}t_A\frac{\partial}{\partial u_i}
\Omega_A(p(\bu),\bu)+\frac{\partial\mathcal{S}_-}{\partial u_i},
\end{equation}
and we may rewrite
\begin{align}\label{31}
\nonumber \Omega_{(i,0)}&=\ln\frac{1}{z-z_{0,i}}-P_{(i,-)}
\Big(\Omega_{(i,0)}-\frac{1}{z-z_{0,i}}\Big),\quad 2\leq i\leq
M\\\\
\nonumber \Omega_{(\alpha,n)}&=z_{\alpha}^n-P_{(\alpha,-)}
\Big(\Omega_{(\alpha,n)}-z_{\alpha}^n\Big),\quad n\geq 1,
\end{align}
where $P_{(\alpha,-)}:=1-P_{(\alpha,+)}$ are the projectors which
annihilate the singular terms of Laurent expansions at
$p=q_{\alpha}(\bu)$. Thus, by noticing that the first terms in the
right-hand sides of \eqref{31} are $\bu$-independent while the
second terms are analytic at $q_{\alpha}(\bu)$, we conclude that
the functions \eqref{27} are also analytic at the points
$q_{\alpha}(\bu)$. Hence, these functions are analytic on the
whole domain $\cal{D}$. Moreover, by taking \eqref{21} into
account, it follows that there is an expansion of the form
\begin{equation}\label{32}
\frac{\partial}{\partial
u_i}\mathcal{S}(p(z,\bu),\bu,\bt)=\sum_{n=1}^{\infty}\frac{s_{i,n}(\bu,\bt)}{p^n}.
\end{equation}
so that
\begin{align*}
\frac{\partial}{\partial
u_i}\mathcal{S}(p(z,\bu),\bu,\bt)&=P_{(1,-)}\Big(\frac{\partial}{\partial
u_i}\mathcal{S}(p(z,\bu),\bu,\bt)\Big)\\
&=P_{(1,-)}\Big(\frac{\partial \mathcal{S}}{\partial
p}R_i\Big)+\frac{\partial \mathcal{S}_-}{\partial u_i}.
\end{align*}
Let us now denote by $E=E(p,\bu)$ any entire function of $p$ such
that
\[
E(p_i(\bu),\bu)=F_i(\bu),\quad i=1,\ldots,N.
\]
Then, by taking into account that \eqref{24} implies
\[
P_{(1,-)}\Big(\Big( \frac{\partial \mathcal{S}_+}{\partial
p}+E\Big)R_i\Big)=0,
\]
it follows that
\begin{align*}
\frac{\partial}{\partial u_i}\mathcal{S}(p(z,\bu),\bu,\bt)&=
P_{(1,-)}\Big(\frac{\partial \mathcal{S_-}}{\partial
p}R_i-ER_i\Big)+\frac{\partial \mathcal{S}_-}{\partial u_i}\\
&=\frac{\partial \mathcal{S}_-}{\partial p}R_i+\frac{\partial
\mathcal{S}_-}{\partial u_i}-\sum_k \frac{r_{ik}F_k}{p-p_k}\\&=0.
\end{align*}
Hence, the statement follows.
\end{proof}

\subsection{Diagonal reductions, symmetric conjugate nets and potentials}

 In the case of diagonal reductions $r_{ij}=\delta_{ij}r_i$,
\begin{equation}\label{42}
\frac{\partial p}{\partial u_i}=-\frac{r_i(\bu)}{p-p_i(\bu)},
\end{equation}
with $i=1,\dots,N$, the compatibility conditions \eqref{16} and
\eqref{23} reduce to
\begin{equation}\label{43}
\begin{aligned}
 \dfrac{\partial r_i }{\partial u_j}
 &=2\frac{r_ir_j}{(p_j-p_i)^2},\\
  \dfrac{\partial p_i }{\partial u_j}&=\frac{r_j}{p_j-p_i}, \\
   \dfrac{\partial F_i }{\partial u_j}&=r_j\frac{F_j-F_i}{(p_j-p_i)^2},
\end{aligned}
\end{equation}
where $i\neq j$. We may extend our observations of \cite{19} by
showing that the diagonal reductions of the Whitham hierarchy
determine a particular symmetric conjugate net as well as a set of
$(M+1)$ Comberscure transformed symmetric conjugate nets. In
particular we are going to prove that the coefficients
$a_{1,1},a_{1,2},a_{1,3},a_{\alpha,-1}$ and $q_\alpha$ are
geometrical potentials associated with these Comberscure
transformed nets.


A conjugate net with curvilinear coordinates $\bu$  can be
described in terms of a set
of rotation coefficients $\{\beta_{ij}(\bu)\}_{\substack{i,j=1,\dots,N\\
i\neq j}}$ which satisfy the Darboux equations \cite{22}
\[
\frac{\partial \beta _{ij}}{\partial u_k}=\beta_{ik} \beta_{kj}
\]
for any triple of different labels $i,j,k$. The associated
Lam{\'e} coefficients $\{H_i(\bu)\}_{i=1,\ldots,N}$ are defined by
the solutions of the linear system
\[
\frac{\partial H _i}{\partial u_j}=\beta_{ji}H_j.
\]

Under a Comberscure transformation a conjugate net transforms into
a parallel conjugate net. The rotation coefficients are left
invariant but the Lam{\'e} coefficients change.  The new Lam{\'e}
coefficients are given by
\[\tilde H_i=\sigma_i H_i
\]
with
\[
\dfrac{\partial \sigma_i }{\partial
u_j}=\beta_{ji}\dfrac{H_j}{H_i}(\sigma_j-\sigma_i).\]

A conjugate net is said symmetric iff $\beta_{ij}=\beta_{ji}$.
Given any pair of parallel symmetric conjugate nets characterized
by $\{\beta _{ij},H_j\}$ and $\{\beta _{ij},\tilde H_j\}$,
respectively; then, it follows that locally there exists a
potential function $\rho$ so that $\sigma _iH_i^2=\dfrac{\partial
\rho}{\partial u_i}$; to see this just observe that
\[
\frac{\partial H_i\tilde H_i}{\partial u_j}=\beta_{ij}(H_i\tilde
H_j+H_j\tilde H_i),
\]
which is a symmetric expression provided $\beta_{ij}=\beta_{ji}$.

Taking $H_i:=\sqrt{r_i}$ and
$\beta_{ij}:=\dfrac{\sqrt{r_ir_j}}{(p_i-p_j)^2}$, as the first
equation on \eqref{43} is $\dfrac{\partial H_i }{\partial
u_j}=\beta_{ij}H_j$, we can identify $H_i$ and $\beta_{ij}$ as the
Lam{\'e} and rotation coefficients, respectively, of a conjugate
net.

The functions $\dfrac{\d \Omega_{i,n}}{\d p}\Big|_{p=q_\alpha }$
determining the hodograph relations are  polynomials in
\[
p_{i,\alpha }=\begin{cases}p_i, &\text{for $j=1$},\\
\dfrac{1}{p_i-q_\alpha },& \text{for $\alpha =2,\dots M$};
\end{cases}
\]
observing that $\beta_{ij}H_j/H_i=\dfrac{r_j}{(p_i-p_j)^2}$ it is
easy to see that these coefficients determine a set of $M$
Comberscure transformations. Then, together with the set of
Lam{\'e} coefficients $\{H_i=\sqrt{r_{i}}\}_{i=1}^N$ we have the
$M$ families of Lam{\'e} coefficients
\[
\{H_{i,\alpha }:=p_{i,\alpha }\sqrt{r_i}\}_{i=1}^N,\quad \text{for
$\alpha =1,\dots, M$}.
\]
 It also follows that there is another
 Comberscure transformed net with Lam{\'e} coefficients
given by
\[
\{h_i:=\sqrt{r_i}F_i\}_{i=1}^N.\]

From \eqref{poten0} and \eqref{poten1}
 we easily find the potentials for
$H_iH_{i,j}$ and $H_{i,j}^2$:
\begin{align}\label{h1}
H_i^2&=-\frac{\partial a_{1,1}}{\partial u_i},\\
\label{h2}H_iH_{i,\alpha }&=\begin{cases}-\dfrac{\partial
a_{1,2}}{\partial u_i}&
 \text{for $\alpha =1$}\\
 -\dfrac{\partial q_\alpha }{\partial u_i} &\text{for $\alpha =2,\dots,M$},
\end{cases}\\
\label{h3}H_{i,\alpha }^2&=
\begin{cases}-\dfrac{\partial (a_{1,3}+a_{1,1}^2/2)}{\partial u_i}&
 \text{for $\alpha =1$},\\
 -\dfrac{\partial\log a_{\alpha ,-1}}{\partial
u_i}&\text{for $\alpha =2,\dots,M$}.\end{cases}
\end{align}
In this way $a_{1,1},a_{1,2},a_{1,3},a_{\alpha ,-1}$ and $q_\alpha
$, $\alpha =2,\dots,M$ acquire a direct geometrical meaning.

Observing that
\[
\beta_{ij}=\dfrac{\sqrt{r_ir_j}}{(p_{i,\alpha }-p_{j,\alpha
})^2}p_{i,k\alpha }^2p_{j,\alpha }^2,\quad \text{for $\alpha
=2,\dots,M$},
\]
we  write our original compatibility conditions as follows
\begin{equation}\label{43-2}
\begin{aligned}
 \dfrac{\partial r_i }{\partial u_j}
 &=2\frac{r_ir_j}{(p_{j,\alpha }-p_{i,\alpha })^2}p_{i,\alpha }^2p_{j,\alpha }^2,\\
  \dfrac{\partial p_{i,\alpha } }{\partial u_j}&=\frac{r_j}{p_{j,\alpha }-p_{i,\alpha }}
  p_{i,\alpha }^2p_{j,\alpha }^2, \\
   \dfrac{\partial F_i }{\partial u_j}&=
   r_j\frac{F_j-F_i}{(p_{j,\alpha }-p_{i,\alpha })^2}p_{i,\alpha }^2p_{j,\alpha }^2,
\end{aligned}
\end{equation}
for $\alpha =2,\dots,M$. This system determines a particular
symmetric conjugate net and two Comberscure transformations of it.
Moreover, if we want to recover the original formulation from
these $p_{i,\alpha }$ we just need the potential $q_\alpha $ of
$p_{i,\alpha }r_i$ and then $r_i,p_i=p_{i,\alpha }^{-1}+q_\alpha
$, $\alpha =2,\dots,M$, will fulfill \eqref{43}.

From \eqref{h1}, \eqref{h2} and \eqref{h3}  we easily get
\begin{align*}
&\frac{\partial^2 a_{1,1}}{\partial u_i\partial u_j}+
\beta_{ji}\sqrt{r_i}\sqrt{r_j}=0,\\
&\begin{cases}-\dfrac{\partial a_{1,2}}{\partial u_i\partial
u_j}+\beta_{ji}\sqrt{r_i}\sqrt{r_j}(p_i+p_j)=0& \\
 \dfrac{\partial^2 q_\alpha }{\partial u_i\partial u_j}+
 \beta_{ji}\sqrt{r_i}\sqrt{r_j}(p_{i,\alpha }+p_{j,\alpha }) =0,&\text{for $\alpha =2,\dots,M$},
\end{cases}\\
&\begin{cases}\dfrac{\partial^2 (a_{1,3}+a_{1,1}^2/2)}{\partial
u_i\partial u_j}+2\beta_{ji}\sqrt{r_i}\sqrt{r_j}p_ip_j=0&
\\
 \dfrac{\partial^2\log a_{\alpha ,-1}}{\partial
u_i\partial u_j}+2\beta_{ji}\sqrt{r_i}\sqrt{r_j}p_{i,\alpha
}p_{j,\alpha }=0&\text{for $\alpha =2,\dots,M$}.\end{cases}
\end{align*}

Observe that \eqref{43} or \eqref{43-2} can be written in terms of
two potentials only. For example we can choose these potentials to
be $q_\alpha $ and $\log a_{\alpha ,-1}$ and use
\[
r_i=-\dfrac{\Big(\dfrac{\partial q_\alpha }{\partial
u_i}\Big)^2}{\dfrac{\partial\log a_{\alpha ,-1}}{\partial
u_i}},\quad p_{i,\alpha }=\dfrac{\dfrac{\partial \log a_{\alpha
,-1}}{\partial u_i}}{\dfrac{\partial q_\alpha }{\partial u_i}}
\]
together with
\begin{align*}
\beta_{ij}\sqrt{r_i}\sqrt{r_j}&=
\dfrac{r_ir_j}{(p_{i,\alpha}-p_{j,\alpha})^2}p_{i,\alpha}^2p_{j,\alpha}^2\\&=
\dfrac{\partial \log a_{\alpha,-1}}{\partial u_i}\dfrac{\partial
\log a_{\alpha,-1}}{\partial u_j}\Big(\dfrac{\partial
q_\alpha}{\partial u_i}\dfrac{\partial q_\alpha}{\partial
u_j}\Big)^2\frac{1}{ W_{ij}^-(a_{\alpha,-1},q_\alpha)^2}
\end{align*}
where \[ W_{ij}^{\pm}(f,g):=\dfrac{\partial f}{\partial
u_i}\dfrac{\partial g}{\partial u_j}\pm\dfrac{\partial a}{\partial
u_j}\dfrac{\partial g}{\partial u_j}
\] for $\alpha=2,\dots,M$.

\begin{gather*}
W_{ij}^-(a_{\alpha,-1},q_\alpha)^2\dfrac{\partial^2q_\alpha}{\partial
u_i\partial u_j}+\dfrac{\partial \log a_{\alpha,-1}}{\partial
u_i}\dfrac{\partial \log a_{\alpha,-1}}{\partial
u_j}\dfrac{\partial q_\alpha}{\partial u_i}\dfrac{\partial
q_\alpha}{\partial u_j}W_{ij}^+(a_{\alpha,-1},q_\alpha)=0,\\
W_{ij}^-(a_{\alpha,-1},q_\alpha)^2\dfrac{\partial^2\log
a_{\alpha,-1}}{\partial u_i\partial u_j}+2\Big(\dfrac{\partial
\log a_{\alpha,-1}}{\partial u_i}\dfrac{\partial \log
a_{\alpha,-1}}{\partial u_j}\Big)^2\dfrac{\partial
q_\alpha}{\partial u_i}\dfrac{\partial q_\alpha}{\partial u_j}=0.
\end{gather*}
For $i,j=1,\dots N$ and $i\neq j$.


\section{Examples}

\subsection{Dispersionless Toda equation}

In order to find solutions of the dT equation
\[
\Phi_{xy}+(\exp(\Phi))_{tt}=0,
\]
we set all $t_A$ equal to zero with the exception of $t_{(2,1)}$
and $t_{(2,0)}$, so that from \eqref{7} and by denoting
\begin{equation}\label{33'}
q(\bt):=q_2(\bt),\quad \nu(\bt):=a_{2,-1}(\bt),
\end{equation}
we have that
\begin{equation}\label{33''}
\Phi=\ln \nu(\bt).
\end{equation}

\paragraph{$N=1$ reduction} Let us first consider reductions $z=z(p,u)$ depending on a
single variable $u$ defined by  $u=-a_{1,-1}$. Then \eqref{14}
becomes the Abel's equation
\begin{equation}\label{33}
\frac{\partial p}{\partial u}=\frac{1}{p-p_1(u)},
\end{equation}
and \eqref{24} reads
\begin{equation}\label{34}
\frac{t}{p_1(u)-q(u)}-\frac{y\,\nu(u)}{(p_1(u)-q(u))^2}+x+F(u)=0,
\end{equation}
where $q(u)$, $p_1(u)$ and $F(u)$ are arbitrary functions of $u$.
On the other hand,
\begin{align*}
\frac{\partial q(u)}{\partial u}&=\frac{1}{q(u)-p_1(u)},\\
\frac{\d \ln\nu}{\d u}&=-\frac{1}{(q(u)-p_1(u))^2}.
\end{align*}

In this way, we may rewrite \eqref{34} as
\[
 t\sqrt{-\frac{\nu'}{\nu}}-y\nu'-x+F(u)=0,
\]
where $\nu':=\d \nu/\d u$. Therefore, as $p_1(u)$ is an arbitrary
function of $u$ we have
\begin{align}\label{35}
\nonumber &tT(u)+yY(u)+xX(u)+F(u)=0,\\ \\
\nonumber &\Phi=\ln\Big(-\frac{XY}{T^2}\Big),
\end{align}
where $T(u)$, $X(u)$, $Y(u)$ and $F(u)$ are arbitrary functions of
$u$. For example when $T,X,Y$ and $F$ are polynomials of 4th
degree we can get explicit examples of solutions. For example,
assuming 2nd order polynomials we get
\[
u:=\gamma\pm\sqrt{\gamma^2-\delta},\quad
\gamma:=\frac{1}{2}\dfrac{X_{1}x+Y_{1}y+T_{1}t+F_{1}}{X_{2}x+Y_{2}y+T_{2}t+F_{2}},\quad
\delta:=\dfrac{X_{0}x+Y_{0}y+T_{0}t+F_{0}}{X_{2}x+Y_{2}y+T_{2}t+F_{2}},\quad
\]
and a solution of dT is
\begin{align*}
\Phi=&\ln\Big((X_1-\gamma
X_2)(-\gamma\pm\sqrt{\gamma^2-\delta})+X_0-\delta X_1\Big)\\&+
\ln\Big((Y_1-\gamma
Y_2)(-\gamma\pm\sqrt{\gamma^2-\delta})+Y_0-\delta Y_1\Big)\\&
-2\ln\Big(-(T_1-\gamma
T_2)(-\gamma\pm\sqrt{\gamma^2-\delta})-T_0+\delta T_1\Big)
\end{align*}

Another example is to take $T=u^3$, $Y=u^2$, $X=u$, $F=1$ to get
the following hodograph relation
\[
t u^3+y u^2+xu+1=0
\]
and the corresponding solution of the dispersionless Toda
equation:
\[
\Phi=3\log\Big(\frac{6tf}{12xt-4y^2+8yf-f^{2}}\Big)
\]
where
\[
f(x,y,t):=\sqrt[3]{36xyt-108t^2-8y^3+12
\sqrt{3}t\sqrt{4x^3t-x^2y^2-18xyt+27t^2+4y^3}}
\]



\paragraph{$N\geq 2$ diagonal reductions}
 Let us consider now reductions $z=z(p,\bu)$ involving $N>1$
variables $\bu:=(u_1,\ldots,u_N)$ associated with a system of
equations \eqref{13} ( or \eqref{14}). Consequently, the functions
$r_{ij}(\bu),\; p_i(\bu)$ are assumed to satisfy the compatibility
conditions \eqref{23}. In this case we
 get the following
system of equations for determining $q(\bu)$ and $\nu(\bu)$
\begin{equation}\label{40}
\begin{aligned}
\frac{\partial q}{\partial u_i}&=R_i(q,\bu),\\
\frac{\partial \ln\nu}{\partial u_i}&=\frac{\partial R_i}{\partial
p}(q(\bu),\bu),
\end{aligned}
\end{equation}
where $i=1,\dots,N$. Thus, given a set of functions
$\{F_i(\bu)\}_{i=1}^N$ satisfying \eqref{23}, the hodograph
relations \eqref{24} read

\begin{align}\label{41}
\nonumber &\frac{t}{p_i(\bu)-q(\bu)}-\frac{y\,\exp\Big(
\int^{\bu}\sum_{j=1}^N \frac{\partial R_j }{\partial p}
(q(\bu),\bu)\d u_j\Big)}{(p_i(\bu)-q(\bu))^2}+x+F_i=0,
\end{align}
for $i=1,\dots,N$,
where now $q$ solves
\begin{equation*}
\frac{\partial q}{\partial u_i}=\frac{r_i}{p_i-q},
\end{equation*}

If we define
\[
P_i:=\dfrac{1}{p_i-q}
\]
the above equation reads
\begin{align}\label{45} \nonumber
tP_i&-yP_i^2\exp\Big( -\int^{\bu}\sum_{j=1}^N r_j P_j^2\d u_j\Big)
+x+F_i=0.
\end{align}

For $N=2$ a simple solution we can take
\begin{equation}\label{47}
\begin{gathered}
r_1=-r_2=\frac{1}{8}(u_1-u_2),
p_1=\frac{1}{4}(3u_1+u_2),\;p_2=\frac{1}{4}(u_1+3u_2),\\\\
F_1=-F_2=\frac{c}{u_2-u_1},
\end{gathered}
\end{equation}
where $c$ is an arbitrary complex constant. In this case we can
get the explicit solution $z(p,\bu)$ of \eqref{13} satisfying
\eqref{12}. It is given by
\begin{equation}\label{48}
z=p+\frac{(u_1-u_2)^2}{16 p-8(u_1+u_2)}.
\end{equation}
Thus, from \eqref{19} we can set
\begin{equation}\label{49}
q(\bu)=-\frac{1}{2}\sqrt{(u_1+z_0)(u_2+z_0)}+\frac{1}{4}(u_1+u_2-2z_0),
\end{equation}
so that by denoting
\[
U_i:=\sqrt{u_i+z_0},\quad i=1,2,
\]
the hodograph relations  become
\begin{equation}\label{50}
\begin{gathered}
x+\frac{4y}{(U_1U_2)^2}-\frac{c}
{(U_1-U_2)^2}=0,\\\\
4t+\Big(2x-\frac{c}{(U_1-U_2)^2}\Big) (U_1+U_2)^2-c=0,
\end{gathered}
\end{equation}
and we have
\begin{equation}\label{51}
\Phi=\ln\frac{(U_1+U_2)^2}{U_1U_2}.
\end{equation}

The system \eqref{50} reduces to a quartic equation as we shall
show. We first write the system \eqref{50} in terms of
\[
u_\pm:=(U_1\pm U_2)^2,
\]
as follows
\begin{equation*}
\begin{gathered}
x+\frac{64y}{(u_+-u_-)^2}-\frac{c}
{u_-}=0,\\\\
4t+\Big(2x-\frac{c}{u_-}\Big) u_+-c=0.
\end{gathered}
\end{equation*}
By eliminating $u_+=(c-4t)(2xu_--c)^{-1}u_-$ we get
\begin{align*}
x^3u_-^4+(-3c+4t)x^2u_-^3+&(4t^2-8ct+3c^2+64xy)x
u_-^2\\&+(-4t^2+4ct-64xy-c^2)cu_-+16c^2y=0,
\end{align*}
and the associated solution \eqref{51} of the dT equation equation
is then given by
\begin{equation}
\Phi=\log\big(\frac{8t-2c}{2t-c+xu_-}\big).
\end{equation}


\subsection{Generalized gas equation}

We consider now solutions of the generalized gas equation
\begin{equation}\label{gas}\begin{gathered}
a_t+(av)_x=0,\\
v_t+vv_x+w_x=0,\\
w_y+a_x=0.
\end{gathered}
\end{equation}
We set all time variables $t_A$ equal to zero except for
$t_{(2,1)}$ and $t_{(1,2)}$ and use the notation conventions
\eqref{33'}. Then, from \eqref{9} it follows that the dependent
variables are given by
\[
a=\nu(\bt),\; v=q(\bt),\; w=a_{1,1}(\bt).
\]

\paragraph{$N=1$ reductions}Reductions $z=z(p,u)$ depending on a single variable $u$, defined
by  $u=-a_{1,-1}$, lead to the Abel's equation \eqref{33} and to a
hodograph relation \eqref{24} of the form
\begin{equation}\label{53}
-t\,p_1-\frac{y\,\nu(u)}{(p_1(u)-q(u))^2}+x+F(u)=0,
\end{equation}
where $q(u)$, $p_1(u)$ and $F(u)$ are arbitrary functions of $u$.
We may rewrite \eqref{53} as
\begin{equation}\label{hodograph-gas}
t\Big( \frac{1}{P(u)}-\int^u P(u)\d u\Big)-yP^2\exp \Big(-\int^u
P^2\d u \Big)
 +x+F(u)=0,
\end{equation}
where $P:=\partial_u q(u)$ is an arbitrary function of $u$ and
\begin{equation}\label{54}
a=\exp\Big(-\int^u P^2\d u\Big),\quad v=\int^u P\d u,\quad w=-u.
\end{equation}

An equivalent form of \eqref{gas} is
\begin{equation}\label{gas2}\begin{gathered}
a_t+(av)_x=0,\\
(v_t+vv_x)_y-a_{xx}=0.
\end{gathered}
\end{equation}

To prove this fact just consider a solution $(a,v)$ of
\eqref{gas2}, then integrating the second equation with respect to
the $y$ variable we conclude the existence of  a function $f(x,t)$
such that
\[
v_t+vv_x-\int_{y_0}^y a_{xx}\d y+f_x(x,t)=0.
\]
Then, a solution of \eqref{gas} is given by $(a,v,w)$ with
\[
w(x,y,t):=f(x,t)-\int^y_{y_0} a_x(x,y,t)\d y.
\]

Now we will show two reformulations of the previous $N=1$
technique providing us with explicit solutions to \eqref{gas2}.
\begin{enumerate}
\item If we parametrize in terms of $a(u)=\exp(-\int^u P^2(u)\d u)$ 
assuming that $a$ is a solution of the following ODE
\[
\frac{\d a}{\d u}=-\dfrac{1}{a f'(a)^2}
\]
for a given function $f=f(a)$ ($f'(a)=\dfrac{\d f}{\d a}$) from
$\log a=-\int^u P^2(u)\d u$ we have $1/P:=-a f'(a)$ and we get the
following hodograph relation
\begin{equation}\label{hodograp-gas-2}
(a f'(a)+f(a))a f'(a)^2t+y-(x+F(a))a f'(a)^2=0.
\end{equation}
Then, given two arbitrary functions $f$ and $F$, and $a(x,y,t)$ a
solution  with of \eqref{hodograp-gas-2} then $a,v=f(a)$ is a
solution of \eqref{gas2}

For example, if $f=Aa+B$ and $F=-(Ca^3+Da^2+Ea+G)$, with
$A,B,C,D,E$ and $G$ arbitrary constants we get the hodograph
relation
\[
A^2Ca^4+A^2Da^3+A^2(2At+E)a^2 +A^2(Bt-Ax+AG)a+y=0
\] and the solution of \eqref{gas2} is $a,v$ with v$v$ given by
\[
\quad v=Aa+B.
 \]
 If we take $f=a+1$ and $F=a^3$ we get
the following solution
\begin{align*}
a&= \alpha-\frac{3(t-x)-4t^2}{9\alpha}-\frac{2t}{3},\\v&=a+1
\end{align*}
with \begin{multline*}
\alpha:=\Big(12t(t-x)-18y-\frac{32}{3}t^3+2\big(12t^3-36xt^2-12t^4+
36tx^2+24xt^3-12x^3\\-12x^2t^2-108t^2y+108txy+81y^2+96yt^3\big)^{1/2}\Big)^{1/3}.
\end{multline*}

Another simple example appears when we take $f(a)=\log a$, the
solution to the hodograph equation (for $F=0$) is
\[
a(x,y,t)=\frac{t}{y}W\Big(\frac{y}{t}\exp(x/t-1)\Big)
\]
and
\[
v(x,y,t)=W\Big(\frac{y}{t}\exp(x/t-1)\Big)-1+\frac{x}{t}
\]
where $W$ is the Lambert function defined by
\[
W(z)\exp(W(z))=z.
\]
 \item Alternatively, we can parametrize in terms of
$v=\int^uP\d u$ where $v$ is subject to the ODE $\dfrac{\d v}{\d
u}=-g'(v)/g(v)$. Then, we get from \eqref{hodograph-gas} the
following hodograph relation
\[
-t(g(v)+v g'(v))g(v)-y{g'(v)}^3+g(v)g'(v)+F(v)+g(v)g'(v)=0.
\]
This equation is gotten by takin into account that
$g(v)=f^{-1}(v)$, is the inverse function of $f$. Thus, given two
arbitrary functions $g,F$ and a solution $v(x,y,t)$ to this
hodograph relation we get a solution $a,v$ with $a$ given by
\[
a=g(v),
 \]
of \eqref{gas2}. In particular if $g:=A\mu^2+B\mu+C$ and
$F=D\mu+E$ we get the following hodograph relation
\begin{align*}
&-3A^2t\mu^4+(2A^2x-8A^3y-5ABt+AD)\mu^3\\&+
(3ABx-12A^2By-2(2AC+B^2)t+AE+BD)\mu^2\\&+
((2AC+B^2)x-6AB^2y-3BCt+BE+CD)\mu\\&+BCx-B^3y-C^2t+EC=0
\end{align*}

\end{enumerate}

%

\paragraph{$N\geq 2$ diagonal reductions} Reductions $z=z(p,\bu)$ involving $N>1$ variables
$\bu:=(u_1,\ldots,u_N)$ can be analyzed by  the same scheme as in
the case of the dT equation. They are associated with a system of
equations \eqref{13} ( or \eqref{14}), where the functions
$r_{ij}(\bu),\; p_i(\bu)$ are assumed to verify the compatibility
conditions \eqref{23}. The functions $q(\bu)$ and $\nu(\bu)$ are
determined by solving the system \eqref{40}. Thus, given a set of
functions $\{F_i(\bu)\}_{i=1}^N$ satisfying \eqref{23}, the
hodograph relations \eqref{24} read

\begin{equation}\label{56}
-t\,p_i(\bu)-\frac{y\,\exp\Big( \int^{\bu}\sum_{j=1}^N
\frac{\partial R_j }{\partial p} (q(\bu),\bu)\d
u_j\Big)}{(p_i(\bu)-q(\bu))^2}+x+F_i=0,
\end{equation}
where $1\leq i\leq N$. The dependent variables of the Generalized
Benney system are then given by
\begin{equation}\label{56'}
\begin{gathered}
a=\exp\Big( \int^{\bu}\sum_{j=1}^N \frac{\partial R_j }{\partial
p} (q(\bu),\bu)\d u_j\Big),\quad v=q(\bu),\\
w= \int^{\bu}\sum_{i,j=1}^N
\operatorname{Res}(R_i(p,\bu),p_j(\bu))\d u_j.
\end{gathered}
\end{equation}

In the particular case of the $N=2$ reduction of diagonal type
defined by
\begin{equation}\label{57}
\begin{gathered}
r_1=-r_2=\frac{1}{8}(u_1-u_2),\\\\
p_1=\frac{1}{4}(3u_1+u_2),\;p_2=\frac{1}{4}(u_1+3u_2),
\end{gathered}
\end{equation}
the function $q(\bu)$ is given by \eqref{49}, so that by denoting
\[
U_i:=\sqrt{u_i+z_0},\quad i=1,2,
\]
we have
\begin{equation}\label{58}
\begin{gathered}
q(\bu)=-\frac{1}{2}U_1U_2+\frac{1}{4}(U_1^2+U_2^2-4z_0),\\
p_1(\bu)=\frac{1}{4}(3U_1^2+U_2^2-4z_0),
p_2(\bu)=\frac{1}{4}(U_1^2+3U_2^2-4z_0).
\end{gathered}
\end{equation}

The hodograph relations \eqref{56} reduce to
\begin{equation}\label{59}
\begin{gathered}
-\frac{4y}{U_1^3U_2}-\frac{1}{4}(3U_1^2+U_2^2-4z_0)t+x+F_1=0,\\\\
-\frac{4y}{U_2^3U_1}-\frac{1}{4}(3U_2^2+U_1^2-4z_0)t+x+F_2=0,
\end{gathered}
\end{equation}
and \eqref{56'} implies
\begin{equation}\label{60}
a=\frac{(U_1+U_2)^2}{U_1U_2},\; v=\frac{1}{4}(U_1-U_2)^2-z_0,\;
w=\frac{1}{16}(U_1^2-U_2^2)^2.
\end{equation}
In particular, for $F_1=F_2=0$ one finds the following explicit
solution

\begin{equation}\label{61}
\begin{gathered}
a=\frac{2}{3}\frac{x+z_0t}{(t^2y)^{1/3}}+2\\
v=\frac{x}{3t}-\Big(\frac{y}{t}\Big)^{1/3}-\frac{2}{3}z_0,\\
w=\frac{(x+z_0t)^2}{9t^2}-\Big(\frac{y}{t}\Big)^{2/3}.
\end{gathered}
\end{equation}

\noindent {\bf Acknowledgements}

This work originated during the stay of the authors at the Isaac
Newton Institute for the Mathematical Sciences of the Cambridge
University. The authors are grateful to the organizers of the
programme "Integrable Systems" for the support provided. They also
acknowledge S. P. Tsarev and A. Mikhailov for many useful comments
and conversations.

\end{document}